\documentclass[twoside,fleqn]{ActaStyle}
\usepackage{times,cite}

\usepackage{graphicx,epsfig}    
\usepackage[tbtags]{amsmath}    

\def\authorheading{B.~Tom\'a\v sik}

\def\shorttitle{Strangeness in ultrarelativistic nuclear collisions}

\begin{document}

\pagerange{1}{4}

\title{STRANGENESS IN ULTRARELATIVISTIC NUCLEAR COLLISIONS
\footnote{Presented at 15-th Conference of Czech and Slovak
Physicists, Ko\v{s}ice, Slovakia, September 5-8, 2005}
\footnote{Based on work done together with E E Kolomeitsev}}

\author{
Boris Tom\'a\v sik\email{boris.tomasik@cern.ch}
}
{
Niels Bohr Institutet, Blegdamsvej 17, 2100 Copenhagen \O, Denmark \\
and 
\'Ustav jadern\'e fyziky AV\v CR, 25068 \v Re\v z, Czech Republic
}

\day{August 15, 2005}

\abstract{%
A model for description of the $\sqrt{s_{NN}}$ dependence of 
$\langle K^+ \rangle/\langle \pi^+\rangle$ ratio at 
the CERN SPS and upper AGS energies is proposed. It uses hadronic degrees
of freedom and the amount of produced strangeness is mainly controlled 
by the total lifetime of the fireball. Decreasing lifetime with increasing 
collision energy is conjectured.
}

\pacs{%
25.75.-q, 25.75.Dw, 24.10.Pa
}

Production of strangeness was among the first observables studied in 
ultrarelativistic nuclear collisions in the search for {\em quark-gluon plasma}
(QGP). The transition into plasma phase was expected to facilitate 
strangeness production and lead to its enhanced abundance \cite{kmr}. It is 
namely rather costly to produce strangeness in a hadronic system; the lowest 
threshold which is about 530~MeV above the incoming masses
is realised in the $\pi N\to \Lambda K$ 
channel. On the other hand, production of a $s\bar s$ pair of quarks 
in the QGP requires only some 300~MeV. Thus the production rates
should be bigger in the deconfined (QGP) phase than in the hadronic gas.

A decisive role in strangeness production is also played by the lifespan 
of the fireball. Smaller or larger production rates cause that
it takes longer or shorter time, respectively, to produce the same 
amount of strangeness. 

\begin{figure}[t]
\begin{center}
\includegraphics[width=12cm, height=5.5cm]{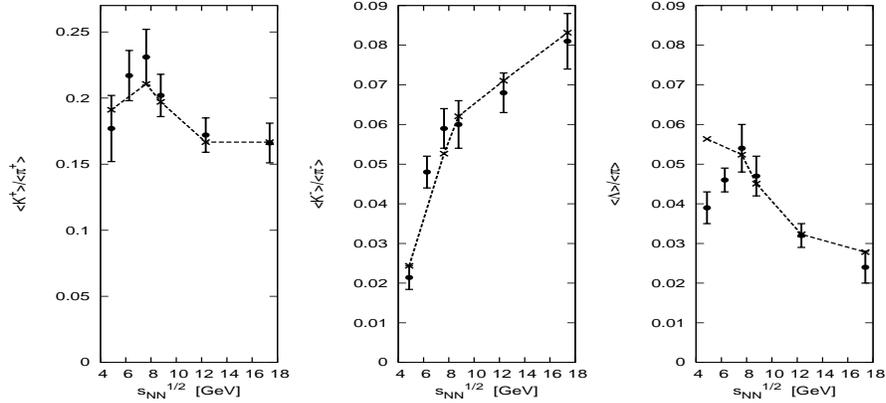} 
\end{center}
\caption{%
Comparison of model calculations for all collision energies with data.
}
\label{f:comp}
\end{figure}

At present, perhaps the most exciting result of the energy scan with 
nuclear collisions at the CERN SPS is the excitation function\footnote{%
Excitation function is, in general, the dependence of a studied quantity 
on the energy of the collision.}
of the $\langle K^+\rangle/\langle \pi^+\rangle$ 
multiplicity ratio (see Figure \ref{f:comp})
\cite{na49data}. So far, the sharp decrease of the ratio above the 
peak position at projectile energy of 30 GeV per nucleon was successfully
interpreted only in framework of a so-called {\em Statistical Model of 
Early Stage} \cite{smes}. This model assumes that primordial\footnote{%
Primordial production is due to collisions of incoming nucleons/partons.}
production of secondaries leads immediately to chemically equilibrated
system where the amount of strangeness is given by the governing temperature
and the phase the system finds itself in. The sharp decrease of the 
$\langle K^+\rangle/\langle \pi^+ \rangle$ 
ratio above $E_{\rm projectile} = 30\,A\mbox{GeV}$  
is put in connection with the mixture of the hadronic and the QGP phase. 

An important task in the hunt for QGP is the exploration of alternative 
models which make {\em no} plasma assumptions. Such models must be falsified
if the identification of QGP is to be claimed. Therefore, we propose
\cite{inprep} a model aimed for interpretation of the $\sqrt{s_{NN}}$
dependence of strangeness production exclusively using hadronic 
degrees of freedom.

It is assumed that the $\langle K^+\rangle/\langle \pi^+\rangle$ peak appears 
due to two effects. Firstly, the rise of positive kaon production
is due to increasing energy which is at disposal for particle 
creation. The second effect is stopping and its weakening toward 
higher collision energies. This leads to shorter lifetimes. As a 
consequence, fewer kaons are produced during a shorter lifespan. 

We focus on densities of various hadronic species and calculate their 
evolution. In particular, we evolve the kaon densities in 
time according to the master equation
\begin{equation}
\label{me}
\frac{dn_K}{dt} = n_K \left ( -\frac{1}{V}\, \frac{dV}{dt}\right ) + 
\sum_{ij} \langle v_{ij}\sigma^+_{ij}  \rangle n_i\, n_j -
\sum_{j} \langle v_{Kj}\sigma^-_{Kj}  \rangle n_K\, n_j
\, .
\end{equation}
The first term on the right-hand side includes the expansion rate. 
It corresponds to density change due to the growth of fireball volume. 
We shall adopt an ansatz for this term expressed via parametrisation 
of the energy density and baryon density as functions of time below in 
eq.~\eqref{tpar}. The second term on the right-hand side of eq.~\eqref{me}
is the kaon production rate; the term in angular brackets denotes 
momentum-averaged cross-sections multiplied by relative velocity 
of the interacting species $i$ and $j$. The last term gives the annihilation
rate and has a similar structure as the second term.

The ansatz for the energy density and densities of conserved quantum 
numbers, $B$ and $I_3$, reads
\begin{equation}
\label{tpar}
\rho(t) = 
\begin{cases}
\rho_0 (1 - at - bt^2)^{\delta} 
& \mbox{for} \quad t < \tau_s \\
\frac{\rho^\prime_0}{(t - \tau_0)^{\alpha\delta}} & \mbox{for} \quad 
t > \tau_s 
\end{cases}
\end{equation}
where $\delta = 1$ for energy density and it assumes a value $3/4<\delta<1$
(depending on the equation of state) for baryon density and $I_3$
density. The $\rho$'s stand for any kind of density here. The first part
of the parametrisation corresponds to acceleration, the second part is a 
power-law expansion suggested by intensity interferometry data
(see e.g.\ \cite{mojrep,lisa} for review). We shall explore a 
range of parameters in this parametrisation and their impact on the results.

From eq.~\eqref{tpar} we determine at any time the energy density and the 
densities of all {\em non-strange} species. These are assumed to be in 
{\em chemical equilibrium} since the reaction rates among them are large. 
Densities of $K^+,\, K^0,\, K^{*+},\, K^{*0}$ are calculated according to 
eq.~\eqref{me}. Negative kaons are of different nature than $K^+$. In a 
baryon-rich environment, as produced in nuclear collisions at these 
energies, the latter are produced in associated production with 
hyperons $\pi N\to Y K$, while the former can only be produced together
with another kaon, e.g.\ $\pi\pi \to K \bar K$. On the other hand, 
reactions which just swap the strange quark from one species to another 
are quick (e.g.\ $\pi \Lambda \leftrightarrow \bar K N$), 
thus the $S<0$ sector 
is in {\em relative} chemical equilibrium: the {\em ratios} of their 
abundances are given by temperature and chemical potentials. Of course,
the {\em total} strangeness of the system must vanish.

In this setup, the final density of $S>0$ species is mainly determined 
by the lifespan and a bit less by the temperature. The total amount of 
negative kaons and $\Lambda$'s is such that the whole $S<0$ sector 
balances the 
$S>0$ densities. The density of $K^-$ {\em relative} to $\Lambda$
is given by the final temperature.

The parametrisation \eqref{tpar} is tuned in such a way that it ends in a 
final state corresponding to chemical composition as obtained in the 
fits by Becattini {\em et al.} \cite{beca}. Primordial content of 
strangeness is estimated from pp, pn, and nn collisions \cite{inprep}.
We are studying the dependence of the resulting density ratios on 
the initial energy density and the total lifetime of the system.

\begin{figure}[t]
\begin{center}
\includegraphics[width=12cm, height=6.5cm]{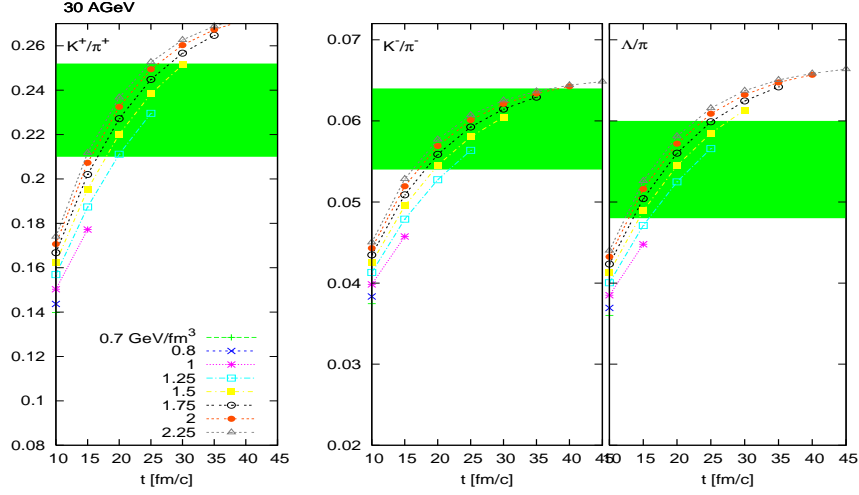} 
\end{center}
\caption{%
Dependence of the multiplicity ratios 
$\langle K^+\rangle /\langle \pi^+\rangle$,
$\langle K^-\rangle /\langle \pi^-\rangle$,
$\langle \Lambda\rangle /\langle \pi\rangle$ on the total lifetime 
of the fireball calculated for Pb+Pb collisions at projectile energy
of 30~$A$GeV. Different curves show scenarios with different initial 
energy densities. Horizontal bands indicate measured values.
}
\label{f:exam}
\end{figure}

An example of the results is shown in Fig.~\ref{f:exam}.
One can observe that strangeness production mainly depends on the lifetime
and not so much on the initial energy density. This feature is yet more
pronounced when increasing the collision energy.
\begin{figure}[t]
\begin{center}
\includegraphics[width=11cm]{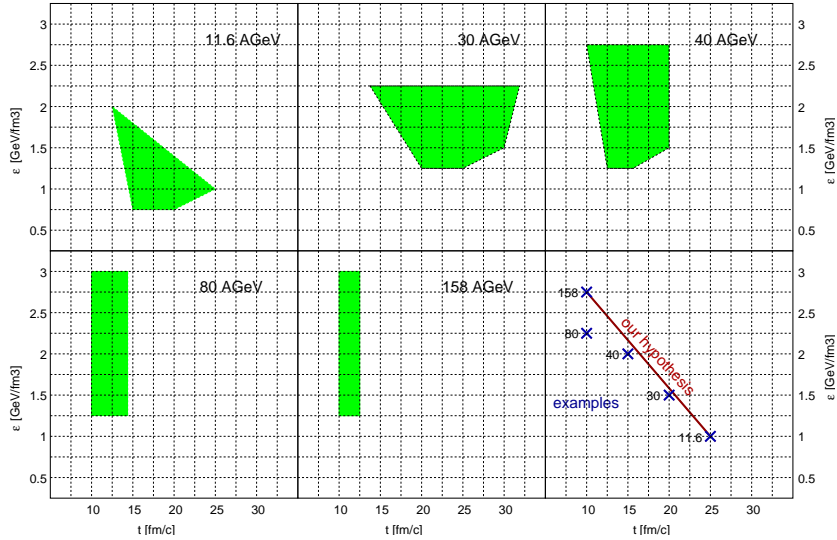} 
\end{center}
\caption{%
Summary of lifespans and initial energy densities allowed by 
comparison with data. Crosses in the lower right panel show the 
parameters of calculations which are compared to data in Fig.~\ref{f:comp}.
}
\label{f:sum}
\end{figure}

In Fig.~\ref{f:sum} we see that at SPS strangeness production is accommodated
by our hypothesis: with increasing collision energy the initial 
energy density is higher and the lifetime shorter. There may be some
problems with collisions at 11.6~$A$GeV (AGS energy); a possible resolution
might be in changing the type of parametrisation for the time-dependence
of densities. The calculated results are compared with data in 
Fig.~\ref{f:comp}.

The underlying idea of the model works. We were able to describe the
data using a hadronic scenario. Whether or not this scenario is
indeed applicable can be decided after a careful cross-check with 
freeze-out analysis of spectra, HBT, abundances, and with the dilepton
spectra.

\begin{ack}
This project was supported by a Marie Curie Intra-European Fellowship
within the 6th European Community Framework Programme.
\end{ack}

\end{document}